\documentclass[11pt,pdftex,a4paper]{article}

\makeatletter
\def\@seccntformat#1{\csname #1ignore\expandafter\endcsname\csname the#1\endcsname\quad}
\let\sectionignore\@gobbletwo
\let\latex@numberline\numberline
\def\numberline#1{\if\relax#1\relax\else\latex@numberline{#1}\fi}
\makeatother

\usepackage[a4paper,top=3.0cm,bottom=2.5cm,left=2.5cm,right=2.5cm]{geometry}

\usepackage{lmodern}
\usepackage[T1]{fontenc} 

\usepackage[greek,english]{babel}
\usepackage{wrapfig}
\usepackage{amsmath}
\usepackage{wrapfig}
\usepackage{graphicx}
\usepackage{sidecap}
\sidecaptionvpos{figure}{c}

\usepackage{bm}
\usepackage{enumerate}
\usepackage{subfigure}
\usepackage{float}
\usepackage{colortbl}
\usepackage{multirow}
\usepackage{rotating}
\usepackage{url}
\usepackage{array}
\usepackage{textcomp}
\usepackage{varioref}
\usepackage{afterpage}
\usepackage{placeins}
\usepackage{fancyhdr}

\usepackage{pdfpages}
\usepackage[small, bf]{caption}

\usepackage{sidecap}
\usepackage[pstricks]{SIunits}
\usepackage[version=3]{mhchem}
\usepackage{tikz}
\usepackage{soul}
\usetikzlibrary{shapes,trees}
\usepackage{wasysym} % diameter symbol

\usepackage{pdflscape} % allows to introduce a landscape page in the document

\usepackage{mathtools} % mathtools contains amsmath

\newcolumntype{C}[1]{>{\centering\arraybackslash }b{#1}}

\usepackage[pdfborder=000,pdftex=true,colorlinks=true,urlcolor=blue,filecolor=black,linkcolor=black,citecolor=black]{hyperref}    
\usepackage{xspace}
\usepackage[figurename=Supplementary Figure]{caption}

\usepackage{upgreek}
\usepackage{placeins}

\begin{document}
	
\renewcommand{\refname}{Supplementary References}



# Quantifying protein densities on cell membranes using super-resolution optical fluctuation imaging

**Tomáš Lukeš**[1,2] **Daniela Glatzová**[3,4], **Zuzana Kvíčalová**[3], **Florian Levet**[5,6], **Aleš Benda**[3,7],**Tomáš Brdička**[4], **Theo Lasser**[1*] & **Marek Cebecauer**[3*]

[1]Laboratoire d'Optique Biomédicale, Ecole Polytechnique Fédérale de Lausanne, Lausanne, Switzerland
[2]Department of Radioelectronics, FEE, Czech Technical University in Prague, Prague, Czech Republic
[3]J. Heyrovsky Institute of Physical Chemistry, Czech Academy of Sciences, Prague, Czech Republic
[4]Institute of Molecular Genetics, Czech Academy of Sciences, Prague, Czech Republic
[5]Interdisciplinary Institute for Neuroscience, UMR 5297 CNRS, Université de Bordeaux, Bordeaux, France
[6]Bordeaux Imaging Center, UMS 3420 CNRS, Université de Bordeaux, US4 INSERM, Bordeaux, France
[7]Imaging methods core facility, BIOCEV, Vestec u Prahy, Czech Republic

*Corresponding authors

**Abstract**

Surface molecules, distributed in diverse patterns and clusters on cell membranes, influence vital functions of living cells. It is therefore important to understand their molecular surface organisation under different physiological and pathological conditions. Here, we present a model-free, quantitative method to determine the distribution of cell surface molecules based on TIRF illumination and super-resolution optical fluctuation imaging (SOFI). This SOFI-based approach is robust towards single emitter multiple-blinking events, high labelling densities and high blinking rates. In SOFI, the molecular density is not based on counting events, but results as an intrinsic property due to the correlation of the intensity fluctuations. The effectiveness and robustness of the method was validated using simulated data, as well as experimental data investigating the impact of palmitoylation on CD4 protein nanoscale distribution in the plasma membrane of resting T cells.

## 1. Introduction

Numerous cellular functions are controlled by molecules at the cell surface among which proteins form the largest pool. An growing body of evidence supports the hypothesis that plasma membrane proteins are not distributed homogeneously but rather in complexes, clusters and other higher order patterns[1]. It has been experimentally demonstrated that these protein clusters are involved in the regulation of signal transduction and other vital cell processes[2]. This has been the driving motivation to develop a robust method for investigation of molecular organization at the plasma membrane under various conditions.

The size of protein assemblies varies and is frequently smaller than 200 nm in diameter which is below the resolution limit of classical fluorescence microscopy. During the last two decades, super-resolution techniques have been developed which overcome the diffraction limit[3,4] and provide a detailed view of structures smaller than 200 nm.

Single molecule localization microscopy (SMLM) has been frequently used to characterize membrane protein assemblies[5–9]. SMLM techniques such as photoactivated localization microscopy (PALM)[10] and stochastic optical reconstruction microscopy (STORM)[11] rely on temporal discrimination of otherwise spatially overlapping fluorophore images.

In sequences of at least several thousand images, the position of fluorescent markers is determined by fitting a model function to the imaged point spread functions (PSFs). In high density samples, this fitting procedure may meet its limit leading to under-counting errors with significant localization errors for overlapping molecules. The stochastic blinking behaviour of of fluorophores may result in multiple localizations from single molecules[12]. High photoswitching rates in combination with high emitter densities can give rise to the appearance of artificial clusters[13]. These limitations may compromise the quantification of densely packed proteins. Characterization of protein clusters becomes a challenge because current methods for cluster analysis[6–9,13–16] rely both on difficult-to-model photophysical properties and on acquisition parameters of the SMLM data. In this work, we readdress these problems with a novel approach based on SOFI and present an innovative and general method

1

to study molecular distribution on cell membranes which overcomes the aforementioned limitations.

SOFI is an optical super-resolution technique which exploits the spatio-temporal photon traces created by stochastically blinking fluorophores. SOFI disentangles the overlapping PSFs by employing higher order statistics. The strong temporal cross-correlation over several neighbouring pixels is the underlying cause of SOFI super-resolution[17,18]. The achieved resolution improvement results from the properties of spatio-temporal cross-cumulants calculated from the entire image sequence of 2D[17] or 3D images[19]. SOFI can be used to analyse SMLM data, but tolerates much higher emitter densities[20,21]. Balanced SOFI (bSOFI) combines the information content of several cumulant orders in a system of linear equations allowing one to extract physically meaningful parameters such as brightness, emitter density and the on-time ratio of the blinking emitters[22]. Therefore molecular density appears as a calculated parameter based on the full image sequence. Multiple blinking of individual fluorophores improves the bSOFI signal and therefore the accuracy of these statistically estimated parameters. In addition, bSOFI suppresses uncorrelated noise[22], which leads to improved image contrast.

We exploited these features of bSOFI and present here a novel SOFI-based quantitative assessment of protein distributions, resulting in protein density maps. In particular, we investigated the impact of palmitoylation on CD4 nanoscale organization at the surface of resting T cells.

## 2. Results

For quantifying the protein distribution in the plasma membrane of T-cells, we acquired image sequences with a total internal reflection fluorescence (TIRF) microscope equipped with an EMCCD camera to detect the fluorescence originating from individual fluorescent emitters (see **Methods**). The proteins of interest were labelled with adequate blinking fluorophores i.e. emitters cycling between dark/bright states.

### 2.1 Molecular density analysis

The algorithm work flow is shown in **Fig. 1**. All acquired image sequences are first drift corrected with sub-pixel precision. Using ThunderSTORM[23], we measured lateral drift using fluorescent beads (fiducial markers) present in the images. These drift corrected image sequences were then processed by our bSOFI algorithm using 2nd, 3rd and 4th order cumulant analysis (see **Methods**). We extracted molecular density maps by combining the cumulant images in a system of linear equations, (see **Methods**). **Fig. 2** shows a data processing example (molecular density analysis) for a single cell. As shown previously[21], the accuracy of the density calculation is mainly determined by the size of the input image sequence. We acquired image sequences of 5000 frames for each dataset, choosing the number of frames by analysing the signal to noise ratio (SNR)[21].

We further evaluated the density maps generated by bSOFI by systematically increasing the density threshold (see **Methods**). Starting with a low threshold, large regions with a low average density are segmented. Increasing the threshold step by step allows precise density quantification (see **Supplementary Fig. 1**). We analysed each region of interest (ROI) by calculating, for each density threshold, the average number and area of high density regions (HDRs), as well as the relative area occupied by the HDRs. The averaged data across all cells for each protein variant over the range of density thresholds is shown in (**Fig. 3a-c**). This analysis provides an overview of HDR parameters in relation to the density threshold, unravelling the overall clustering behaviour of the cell samples under study. Inset images in **Fig. 3a** and **Supplementary Fig. 1** indicate how the density threshold affects the detection of HDRs. Detailed statistics of the quantitative molecular density data can be further presented for the optimal density threshold (**Fig. 3d**; the threshold determined from simulations, see **Supplementary Fig. 2**) or any other threshold selected, for example, based on biological reasoning. When calculating the 4th order SOFI image, the pixel size of the resulting SOFI density map is 26.25 nm. According to the Shannon-Nyquist sampling theorem, the smallest detectable HDR would have a diameter of 52.5 nm. The density analysis can distinguish differences in HDR diameters in increments of 26.25 nm. Higher resolution could be possible with higher order SOFI images at the expense of more input images, i.e. longer acquisition times. The simulations indicate good performance of the analysis across a broad range of HDR densities (500 – 3000 mol/μm) and HDR to background ratios (20 – 100). Accuracy of HDR detection increases with increasing HDR to background ratio (see **Methods** and **Supplementary Fig. 3**).

2

## 2.2 Protein nanoscale organization

We expressed four different mutant variants of mEOS2-labelled CD4 and analysed individual protein distributions on the plasma membrane of resting T cells immobilised on poly-L-lysine coated glass coverslips. Using TIRF microscopy, we imaged 20 cells for each CD4 variant (i.e. 80 in total) acquiring 5000 frames per cell. Tested mutants were native CD4 protein (WT), palmitoylation mutant (CS1) and truncated variants lacking the extracellular (dD1D4) and cytoplasmic (dCT) domains (**Supplementary Fig. 4**). Segmentation of SMLM data acquired for CD4-mEos2(WT) indicated the accumulation of native CD4 in HDRs with irregular shape, frequently forming networks (**Fig. 2** and **Supplementary Fig. 5**). SMLM-based cluster analysis of these localisation data would be a challenge[13]. We therefore based our analysis on bSOFI imaging and extracted the corresponding density maps (see **Methods**). To minimize cell-size dependency and aiming for a true comparative protein density analysis among different CD4 variants, we selected a 3 x 3 µm region of interest (ROI) in each cell. **Fig. 3** summarizes the quantitative data on CD4 membrane organization and indicates significant differences between the tested protein variants at the cell surface of resting T cells. As shown in **Fig. 3d**, native (WT) CD4 are organized in HDRs covering a large part of the plasma membrane as indicated in Roh et al.[24]. Such arrangements depend on the intact extracellular domain and palmitoylation of CD4 since mutants lacking these structures exhibit more random distribution with rare accumulation in a rather small HDRs (**Fig. 3d**). Truncation of the cytoplasmic domain had only a minor effect. The results presented in **Fig. 3** point to the ability of our new method to identify HDRs with irregular shape and varying densities. The imaged cells exhibited a high level of intercellular variability, especially in case of the intermediate phenotype (CD4-dCT), and heterogeneity between HDRs identified within ROIs (see **Supplementary Fig. 6**).

## 3. Discussion

In this work, we introduced a novel method for the characterisation of molecular organisation on cellular surfaces. Our quantitative analysis is based on SOFI, which provides several distinct advantages as applicable to densely populated regions (overlapping fluorescence emitters), no need for multiple blinking corrections, and inherent access to molecular density without *a priori* assumptions about the clustered molecules. Our approach does not require molecular localisation coordinates to calculate clustering properties of proteins (or other molecules) on the cell surface. Our algorithm provides quantitative molecular density analysis of membrane protein distributions independent of any user-defined parameters. We demonstrated the applicability of the proposed method by analysing the surface distribution of CD4 glycoprotein which forms large, dense, and interconnected regions on human T cells. Our molecular density analysis indicates the importance of the extracellular domain and of receptor palmitoylation for the organisation of CD4 on the plasma membrane. Our new method has the potential to be extended for various molecular density studies of surface molecules accessible for fluorescent labelling under physiological, pathological or pharmacological conditions.

# 4. Methods

## 4.1 Microscope setup

A customized setup built on an inverted optical microscope (IX71, Olympus) was used for cell imaging. A 150 mW, 561 nm laser (Sapphire, Coherent) and a 100 mW 405 nm laser (Cube, Coherent) provided the excitation and activation, respectively. An acousto-optic tuneable filter (AOTFnC-400.650-TN, AA Optoelectronics) provided fast switching of the laser sources. Both lasers were combined and focused into the back focal plane of an objective (UApoN 100x, NA=1.49, Olympus). Total internal reflection was achieved with a commercial TIRF module (IX2-RFAEVA-2, Olympus) and the fluorescence emission was detected by an EMCCD camera (iXon DU-897, Andor).

## 1.1 Sample preparation

Jurkat T cells in RPMI-1640 media (Sigma-Aldrich), complemented with glutamine and 10% foetal calf serum (Life Technologies) were grown in an incubator under controlled conditions of 37°C, 5% CO2, and 100% humidity. The cells were transiently transfected using the Neon® transfection system (Life Technologies). 1 µg of vector DNA per shot (3 pulses of 1325 V lasting for 10 ms) per 200,000 cells was used (see manufacturer's instructions). 25 mm diameter microscope coverslips were cleaned by incubation with 2% Hellmanex (Sigma-Aldrich) overnight at 42˚C and subsequently washed with MiliQ water. Prior to use, the coverslips were coated with poly-L-lysine (Sigma-Aldrich). Twenty hours after transfection, the cells were washed with PBS, resuspended in phenol red-free RPMI-1640 media (Sigma-Aldrich), supplemented with 10% foetal calf serum (Life Technologies), seeded on the poly-L-lysine coated coverslips, and incubated for 5 min at 37°C under 5% CO2. After a quick PBS wash the cells were fixed using 4% paraformaldehyde in PBS at 37 °C for 10 minutes under 5% CO2. After removal of excess liquid, the fixation was stopped with 0.1 M NH4Cl in PBS and the cells were washed with PBS. Finally, the coverslip was placed into a ChamLide holder for imaging.

## 1.2 Imaging

Fixed cells were imaged in a 0.9% NaCl solution at room temperature. For monitoring drift, 200 nm gold beads (BBI international) were added to the sample. The mEos2 fluorophore was excited at 561 nm with power of ~30 mW and activated by a 405 nm laser with power of ~ 3 mW (both measured at the sample plane ). Cells were imaged with an EMCCD camera using an EM gain of 300 and an exposure time of 32 ms.

## 4.2 SOFI molecular density analysis

The SOFI molecular density analysis can be subdivided into three distinct steps:

### Drift correction
SOFI needs the sample to be immobile during image acquisition, and imaging beyond the diffraction limit demands drift correction. Tracking the positions of gold nanoparticles provides translational motion vectors in between consecutive frames. Registering consecutive frames with sub-pixel precision using bilinear interpolation was used for drift correction.

### Bleaching correction
The drift corrected image sequence was sub-divided into sub-sequences of 500 frames each. These sub-sequences were chosen sufficiently short in order to minimize the influence of photobleaching[19,21]. In each subsequence, SOFI images of 2nd, 3rd and 4th order were calculated. These SOFI images were then averaged across all sub-sequences.

### Molecular density analysis
The SOFI based molecular density analysis was programmed in MATLAB taking into account a linearization procedure as described in[21]. Combining SOFI images of different orders allows one to extract density maps (see **Supplementary Note**). Molecular density (i.e. number of emitters per pixel area) at pixel position **r** is given as

$$N(\mathbf{r}) = \frac{g_2(\mathbf{r})}{(3K_1^2 - 2K_2) \frac{3K_1^2 - K_1\sqrt{3K_1^2 - 2K_2} - 2K_2}{2(3K_1^2 - 2K_2)} (1 - \frac{3K_1^2 - K_1\sqrt{3K_1^2 - 2K_2} - 2K_2}{2(3K_1^2 - 2K_2)})},$$

where $K_1(\mathbf{r}) = \frac{\mu_2 g_3}{\mu_3 g_2}(\mathbf{r})$, $K_2(\mathbf{r}) = \frac{\mu_2 g_4}{\mu_4 g_2}(\mathbf{r})$, $g_2, g_3, g_4$ represents cumulant images of 2nd, 3rd and 4th order, respectively. $\mu_n = \mathcal{E}_V\{U^n(\mathbf{r})\}$, where $\mathcal{E}_V\{U^n(\mathbf{r})\}$ is the expectation value of the PSF ($U^n(\mathbf{r})$) of the nth order cumulant image. For more details, see **Supplementary Note**.

Areas containing only background are removed using the bSOFI image as a mask. The threshold filtering procedure is described in more detail in **Supplementary Figure 1**. The algorithm loop through a

whole range of density levels presented in the sample starting with a threshold equal to zero and increasing the threshold step by step in each iteration. For each threshold, the data are further processed to acquire quantitative parameters describing regions with local protein density above the threshold i.e. high density regions (HDRs). The algorithm calculates the area, equivalent diameter, number of HDRs, and the proportion of the total area of the ROI covered with HDRs. As a result, we obtain a graph which describes the dependence of HDR parameters on the molecular density and reveals the overall clustering behaviour of the sample under study. The reliability of the algorithm was investigated under a broad range of simulations (see **Supplementary Figure 3**).

The absolute values of SOFI density map depend on expression of the fluorescent markers and parameters of the microscope (particularly excitation intensity and a camera gain). Therefore, we use relative densities and investigate relative changes of local density. For simulations, the molecular density maps were normalized by the mean density calculated over all ROIs. For the experimental data, samples were split into four groups according to four CD4 variants. In each group, a group mean density was calculated across ROIs of twenty samples. ROI of each sample was first normalized to the same mean within the group and then by the maximum of all group means. This normalization procedure largely removes the expression dependence in between the experiments while preserving the differences in relative density in between the CD4 variants. Therefore, normalized SOFI density values can be compared across experiments, as long as the parameters of image acquisition remain the same.

**4.3 Simulations**

The simulation assumed photokinetics as known for fluorescent proteins in PALM experiments[21]. A photon time-trace for each fluorophore was simulated providing the number of emitted photons over time. The pixel intensity at a given time point corresponds to the integration over the brightness originating from fluorophores in the conjugated object localizations. The number of converted photo-electrons was estimated by a Poisson distributed random distribution. The average value was taken as a pixel value multiplied by the detection efficiency. Additive noise corresponding to thermal noise, read out noise and gain variations was added as a Gaussian noise contribution. Optical system and camera parameters are matched to the microscope system settings (NA, wavelength, magnification, pixel size etc.).

The ground truth object is composed of 10 HDRs randomly distributed over an area of 3 x 3 µm. The diameter of generated HDRs varies over the range 60 – 180 nm, whereas the molecular density in HDRs varies over the range 500 – 3000 µm. In between the HDRs, individual molecules were randomly distributed such that the HDR to background ratio was {20, 50, 100}. For each test scenario, we simulated a random distribution of labelled molecules, including no clusters as a control. In total, we generated and analysed 720 simulated image sequences (see **Supplementary Figure 3**). The simulation proves that the algorithm performs well under a broad range of conditions.

## Author contributions

T.Lu., T.La. and M.C. conceived the study. T.Lu., T.La. developed the molecular density analysis. D.G. and Z.K. prepared the samples. D.G, T.Lu. performed the experiments, T.Lu., F.L., and Z.K. analyzed the data. A.B. made the microscope setup. A.B., T.B. provided research advice. T.Lu., T.La. and M.C. wrote the paper. All authors reviewed and approved the manuscript.

5

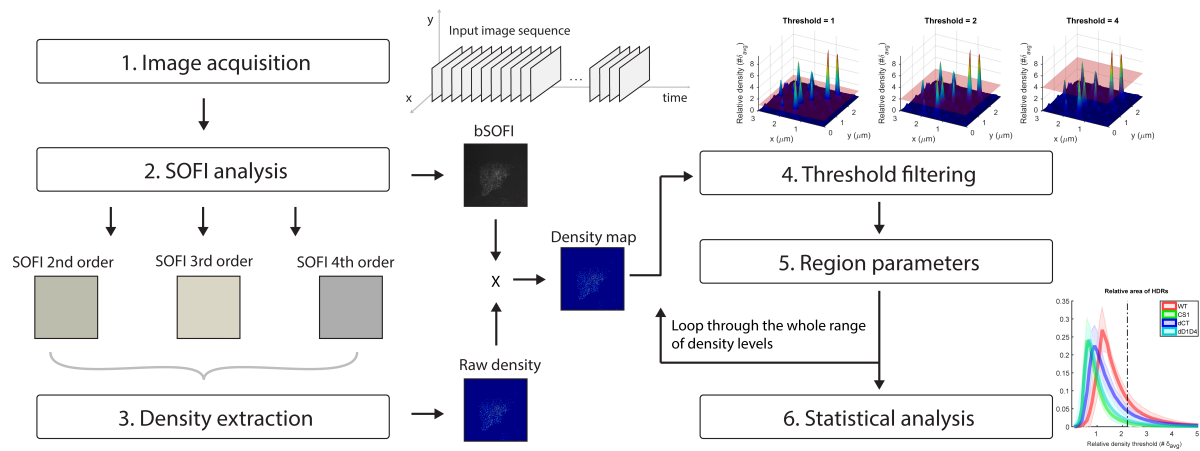

**Figure 1:** The workflow of SOFI-based molecular density analysis. SOFI images of different cumulant orders were calculated and used to extract molecular densities. The background was removed using the bSOFI image as a transparency mask. High density regions (HDRs) were segmented by varying the threshold parameter over the whole range of available density levels. For each threshold, the area, equivalent diameter, and number of HDRs were extracted and plotted as a function of the density threshold. (see **Fig. 3**; given as multiples of mean density in tested ROIs). This procedure is then repeated for each sample and ROI.

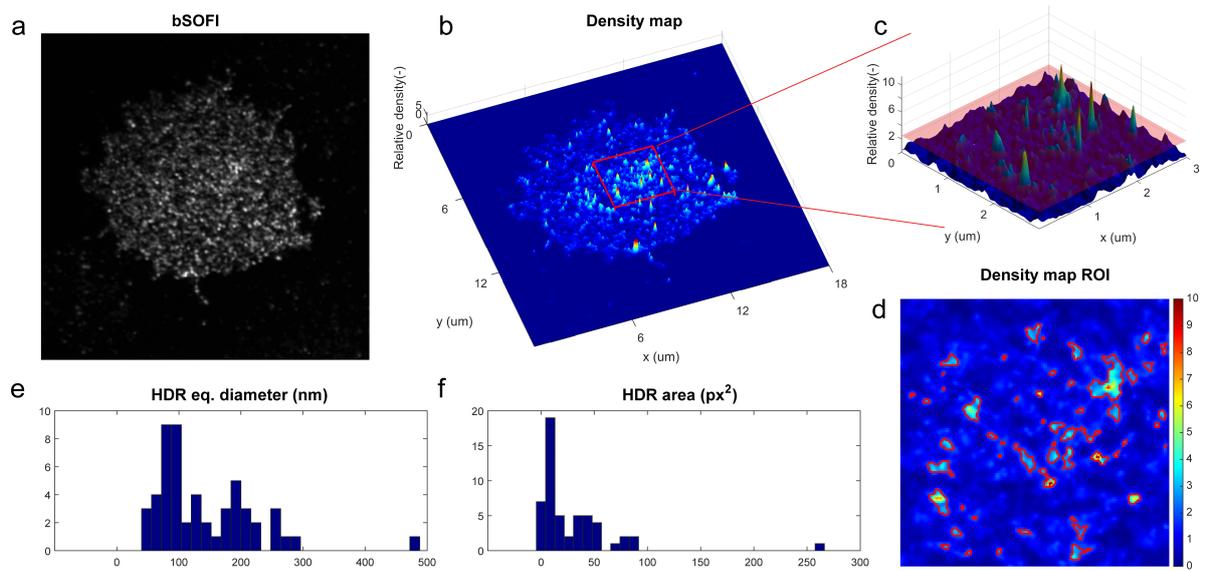

**Figure 2:** Example of data processing for a single cell expressing CD4(WT)-mEos2 fusion protein. (**a**) bSOFI image. (**b**) Molecular density map. (**c** and **d**) Segmentation of the 3 x 3 µm region of interest indicated in **b** by the red square for a relative density threshold equal to 2.2 times the mean density. (**e**) Histogram of equivalent diameters (i.e. diameter of a circle of the same area as the non-circular region). (**f**) Histogram of measured area (in $px^2$) of high density regions (HDRs) in the ROI shown in **d**.

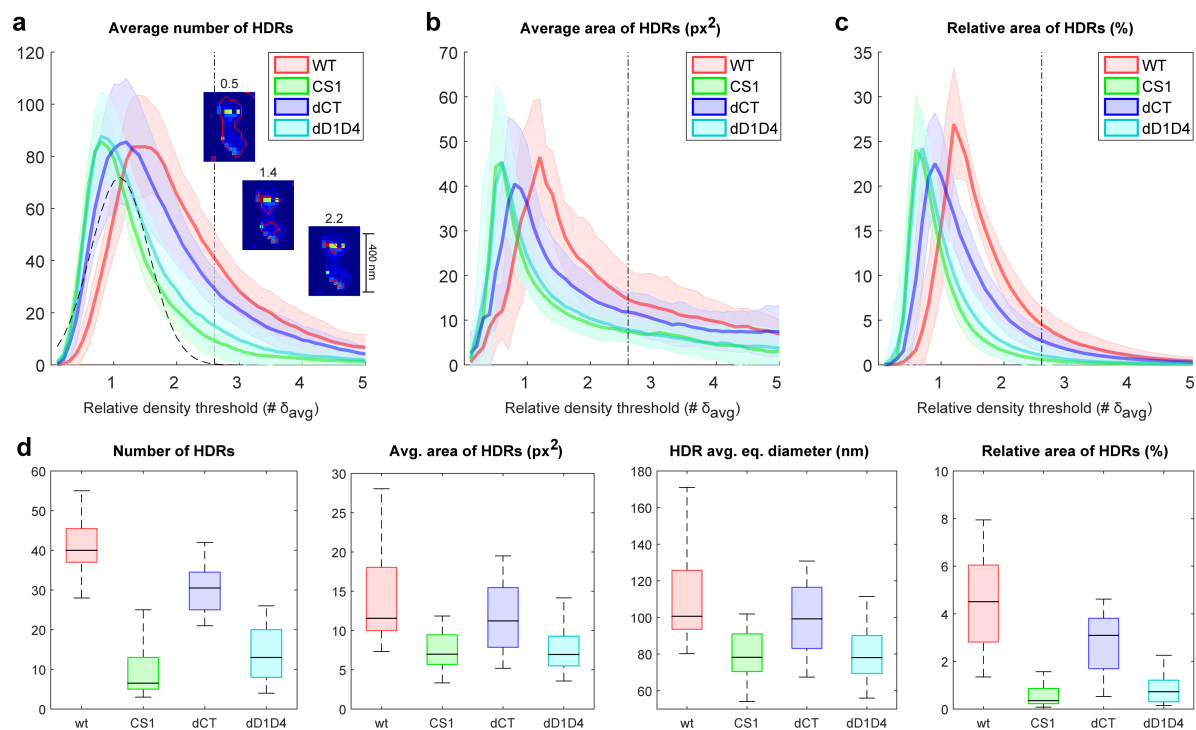

**Figure 3:** SOFI analysis of four CD4 protein variants in resting T cells immobilized on poly-L-lysine coated coverslips. Native CD4 (WT), palmitoylation mutant (CS1) and variants lacking the extracellular (dD1D4) and cytosolic parts (dCT) were tested (n=20 per variant). (**a**) Number of high density regions (HDRs) averaged over all samples for each CD4 variant. Density thresholds are related to the mean density calculated over the 3 x 3 μm ROIs of all samples. The inset images show examples of the segmented HDRs for various thresholds indicated above the image. (**b**) HDR area averaged over all samples for each CD4 variant in $px^2$, where pixel size is 25 nm. (**c**) Relative area occupied by HDRs related to the total area of the ROI. (**d**) Box-plots of the properties of HDRs for a threshold equal to $2.6\delta_{avg}$ (marked by the vertical dash-dot line). The chosen threshold is the value, where Gaussian function of a random distribution (marked in **a** by the dashed line) falls below 1 (see **Supplementary Fig. 2**). Semi-transparent colour areas in **a,b,c** represent standard deviation. In each box-plot in **d**, the box represents the interquartile range (IQR), the central mark is the median, and the whiskers extend to the most extreme data points.

9


	
	%%------------------------------- Header Definition
	
	%\fancyfoot[C]{\thepage} 
	\headheight 15.0pt 
	% redefine plain style
	\fancypagestyle{plain}{% 
		\fancyhf{} % clear all header and footer fields 
		\renewcommand{\headrulewidth}{0pt} 
		\renewcommand{\footrulewidth}{0pt}} 
	\setcounter{tocdepth}{2}
	%%------------------------------------------------
	
	\section*{Supplementary Material for the article:}
	
	\section*{Quantifying protein densities on cell membranes using super-resolution optical fluctuation imaging}

	Tom\'{a}\v{s} Luke\v{s}\textsuperscript{1,2},
	Daniela Glatzov\'{a}\textsuperscript{3,4}, Zuzana Kv\'{i}\v{c}alov\'{a}\textsuperscript{3}, Florian Levet\textsuperscript{5,6}, Ale\v{s} Benda\textsuperscript{3,7},Tom\'{a}\v{s} Brdi\v{c}ka\textsuperscript{4}, Theo Lasser\textsuperscript{1} \& Marek Cebecauer\textsuperscript{3}\\
	\\
	\textsuperscript{1}Laboratoire d'Optique Biom\'{e}dicale, Ecole Polytechnique F\'{e}d\'{e}rale de Lausanne, Lausanne, Switzerland
	\\	
	\textsuperscript{2}Department of Radioelectronics, Faculty of Electrical Engineering, Czech Technical University in Prague, Prague, Czech Republic
	\\
	\textsuperscript{3}J. Heyrovsky Institute of Physical Chemistry, Academy of Sciences, Prague, Czech Republic
	\\
	\textsuperscript{4}Institute of Molecular Genetics, Czech Academy of Sciences, Prague, Czech Republic
	\\
	\textsuperscript{5}Interdisciplinary Institute for Neuroscience, UMR 5297 CNRS, Universit\'{e} de Bordeaux, \\Bordeaux, France
	\\
	\textsuperscript{6}Bordeaux Imaging Center, UMS 3420 CNRS, Universit\'{e} de Bordeaux, US4 INSERM, \\Bordeaux, France
	\\
	\textsuperscript{7}Imaging methods core facility, BIOCEV, Vestec u Prahy, Czech Republic
	\\

\newpage
			
% Supplementary Fig. 1+
\begin{figure}
	\centering\includegraphics[width=1\textwidth]{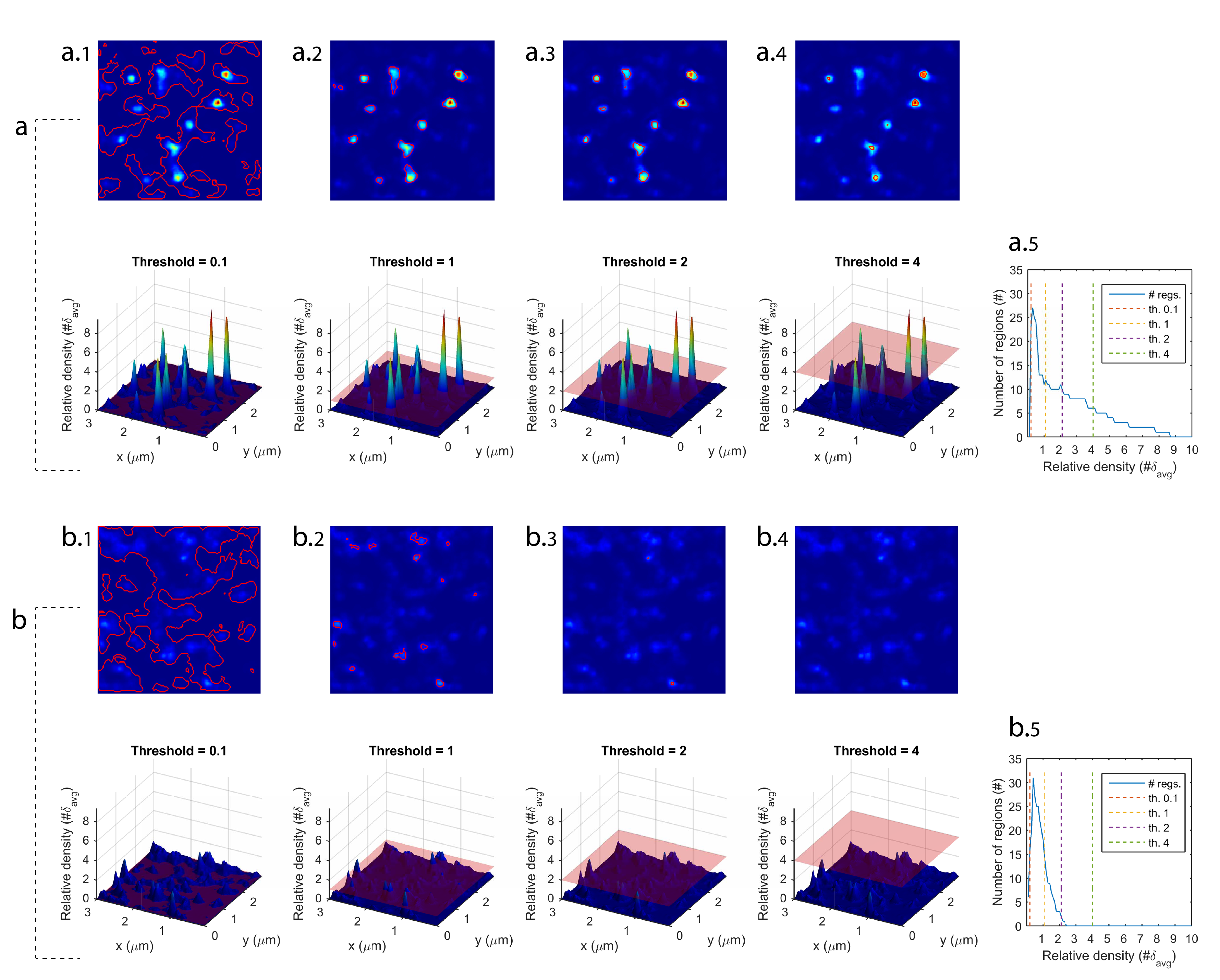}
	\caption{\\
		\\
		SOFI-based molecular density analysis - threshold filtering.\\
		\\
		Threshold filtering of simulated datasets containing (a) high density clusters, (b) randomly distributed emitters. First a mean density over the wall region of interests (ROIs) is calculated as $ \delta_\mathrm{avg} = \frac{1}{KLN}\sum_{k=1}^{K} \sum_{l=1}^{N}\sum_{n=1}^{N}d(k,l,n)$, where $ d(k,l,n) $ is a molecular density per pixel located in the k-th row and l-th column of n-th ROI, N is the total number of ROIs, k,l runs through all rows and columns of the ROI, respectively. The threshold parameter is given as a multiple of the mean density taken over the selected ROI. i.e. $ threshold = 2 $ corresponds to $ 2\delta_\mathrm{avg} $. For each threshold setting, densities above the threshold determine the boundary of the density dependent area providing number of segments, area size and equivalent diameter. (a.1 - a.4) shows an example for threshold values: 0.1, 1, 2, and 4. Repeating this procedure step by step for the whole range of thesholds, we obtain charts  that show number of HDRs as a function of the density threshold (a.5,b.5; blue line) for the case with HDRs (a.5) and with randomly distributed molecules (b.5). For the first case, 6 HDRs are detected at the density threshold 4 (a.5; green dashed line). For the second case, 0 HDRs is detected at this density threshold (b.5; green dashed line).}
	\label{fig:threshSteps} 
\end{figure} 
	
% Supplementary Fig. 2
\begin{figure}
	\centering\includegraphics[width=1\textwidth]{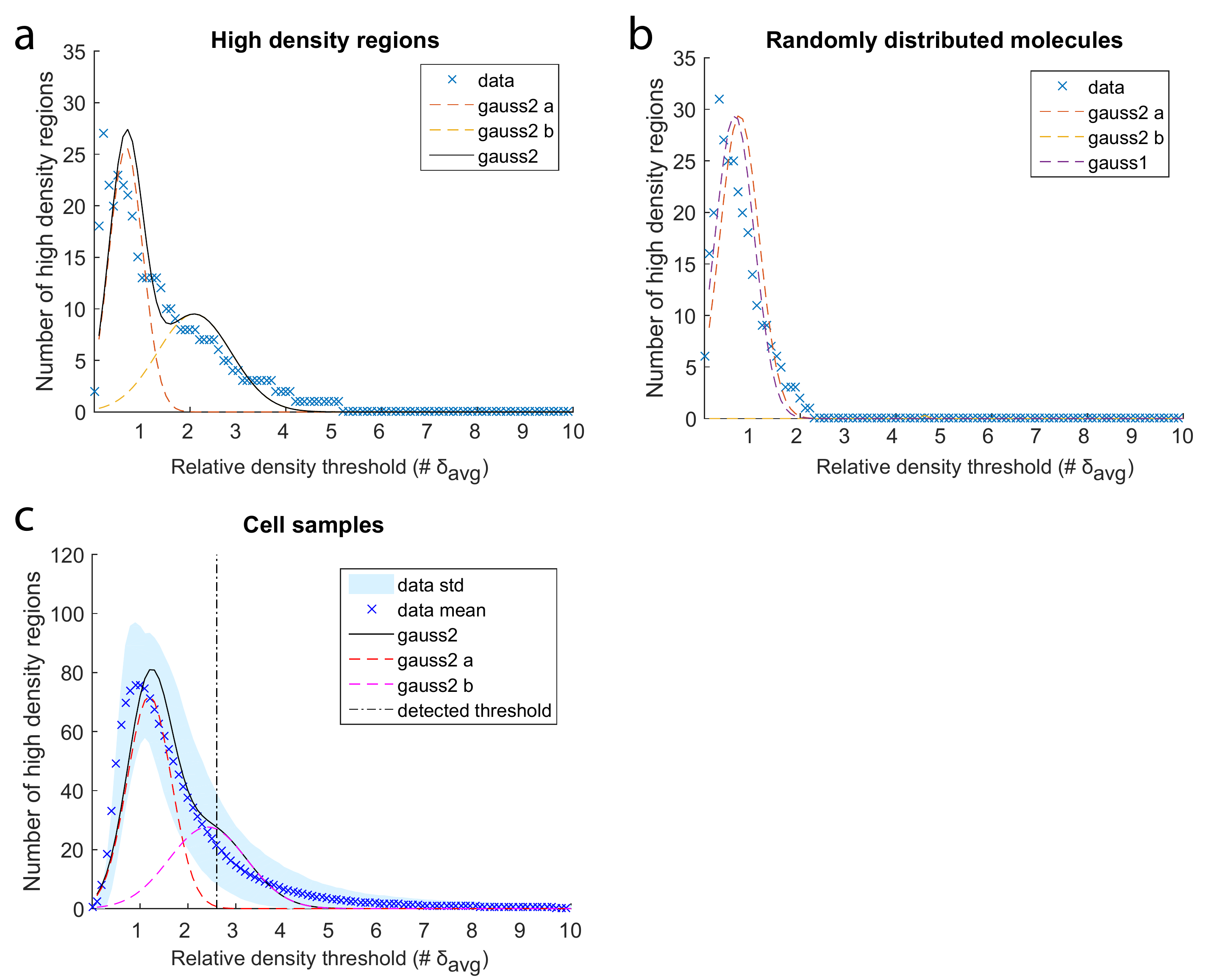}
	\caption{\\
		\\
		SOFI-based molecular density analysis - threshold detection.\\
		\\
		Number of high density regions (HDRs) as a function of density threshold for a simulated sample which (a) contains high density regions, or (b) contains randomly distributed molecules. Fitting a sum of two Gaussian functions reveals a component which corresponds to the random patterns (red dashed line) and a component corresponding to non random HDRs (yellow dashed line). (c) Number of HDRs as a function of density threshold averaged over all cell samples (i.e. 80 samples). Averaging across all samples allows us to obtain one density threshold for all samples and thus compare HDRs size of different CD4 variants at the same density level (Fig. 3). A sum of two Gaussian functions fitted to the data ("gauss2"). Vertical dash-dot line indicates the detected threshold where the value of the Gaussian function (red dashed line), which corresponds to randomly distributed molecules, falls below 1.}
	\label{fig:gaussFitting} 
\end{figure}

\FloatBarrier
% Supplementary Fig. 3
\begin{figure}
	\centering\includegraphics[width=0.94\textwidth]{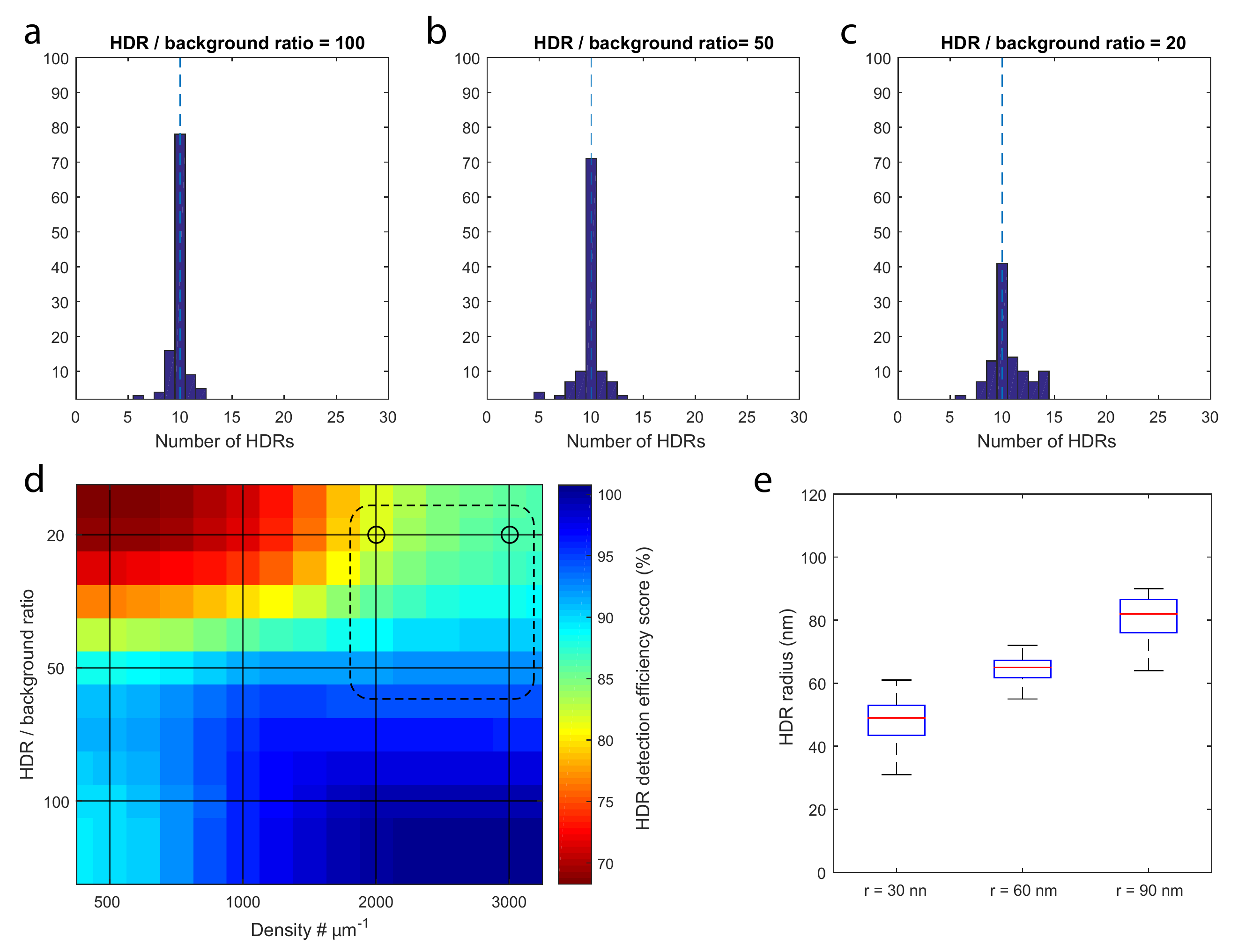}
	\caption{\\
		\\
		Simulation of 	SOFI-based molecular density analysis under controlled conditions.\\
		\\
		Estimation of number of high density regions (HDRs): 
		10 simulated HDR per ROI as ground truth.
		The HDR radius was in the range $ \{30, 60, 90\} $ nm, 
		the molecular density per cluster was in the range $ \{500,1000,2000,3000\} $ molecules per $ \mu m^2  $. 
		In between the HDRs, molecules were randomly distributed such that HDR/background ratio was equal to $ \{100,50,20\} $. Each test scenario was repeated 10 times. In total, 360 datasets were generated and evaluated (120 datasets for each HDR/background case). The number of photons per emitter per frame was set to 100, which corresponds well to the experimental conditions. The number of frames of each image sequence was 5000. Dashed blue line in (a),(b),(c) marks the ground truth. 
		\\
		Overall, the simulation validates the algorithm and estimation under a broad range of conditions.  (d) HDR detection efficiency score is a probability that the estimated number of HDRs is in the range (7-13). The accuracy of the estimation increases with increasing HDR/background ratio and increasing HDR density. The simulations were calculated for the grid points. The dashed line marks the conditions where the signal to noise ratio (SNR) and signal to background ratio (SBR) of simulated data correspond to SNR and SBR of the real cell datasets in our experiments. (e) Estimation of HDR radius for grid points marked in \textbf{d} by the circles. }
	\label{fig:numHistograms} 
\end{figure} 

% Supplementary Fig. 5
\begin{figure}
	\centering\includegraphics[width=0.6\textwidth]{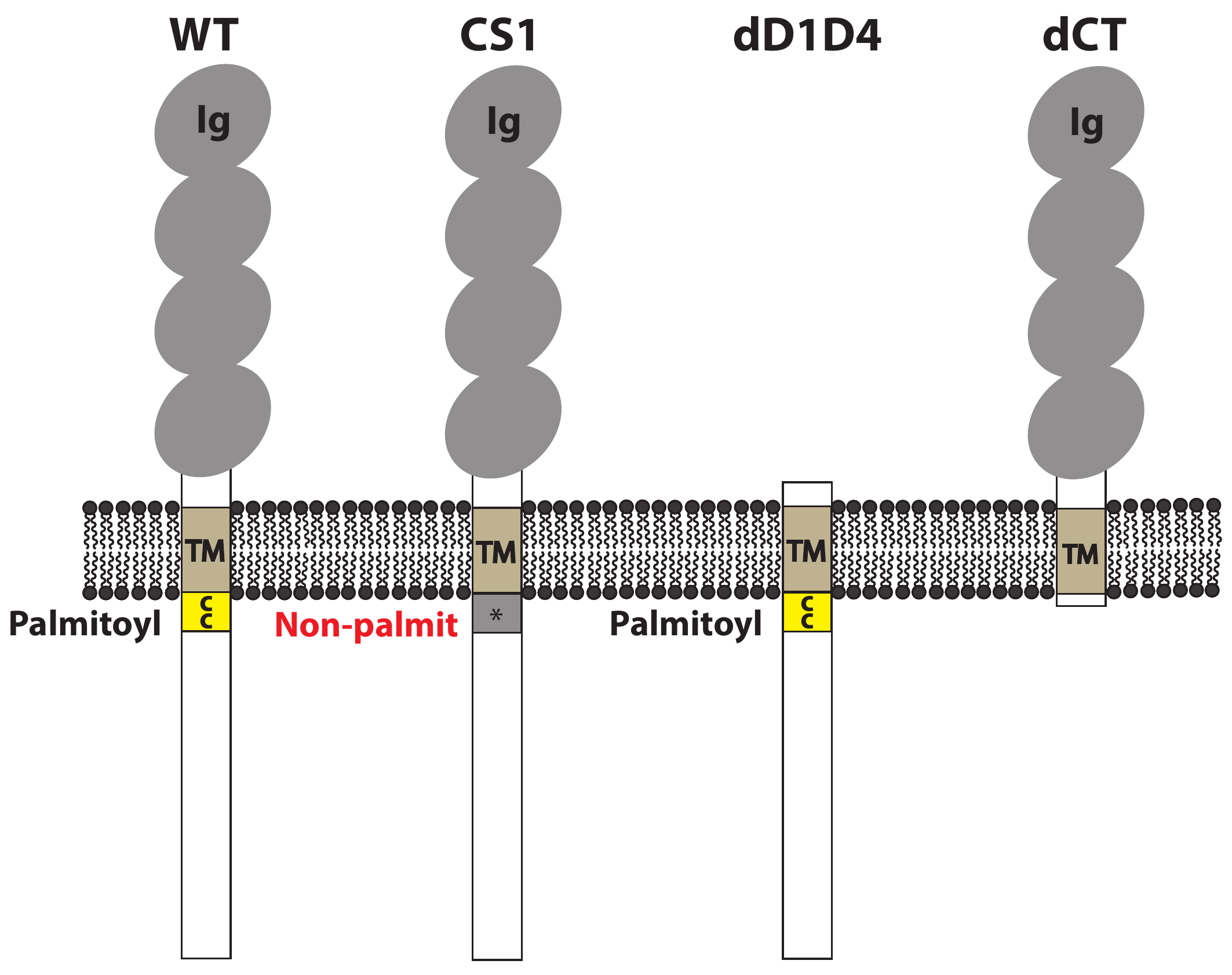}
	\caption{Schematic representation of wild-type (WT) CD4 and its variants: palmitoylation mutant (CS1), mutant missing the extracellular domain (dD1D4) and mutant missing the intracellular domain (dCT). TM = transmembrane domain, Ig = four immunoglobulin type domains D1-D4, Palm = palmitoylations sites C419 and C422, Non-palm = non-palmitoylated variant with mutations C419,422S. Proportions are not in scale.}
	\label{fig:CD4variants} 
\end{figure}

% Supplementary Fig. 6 SMLM examples
\begin{figure}
	\centering\includegraphics[width=1\textwidth]{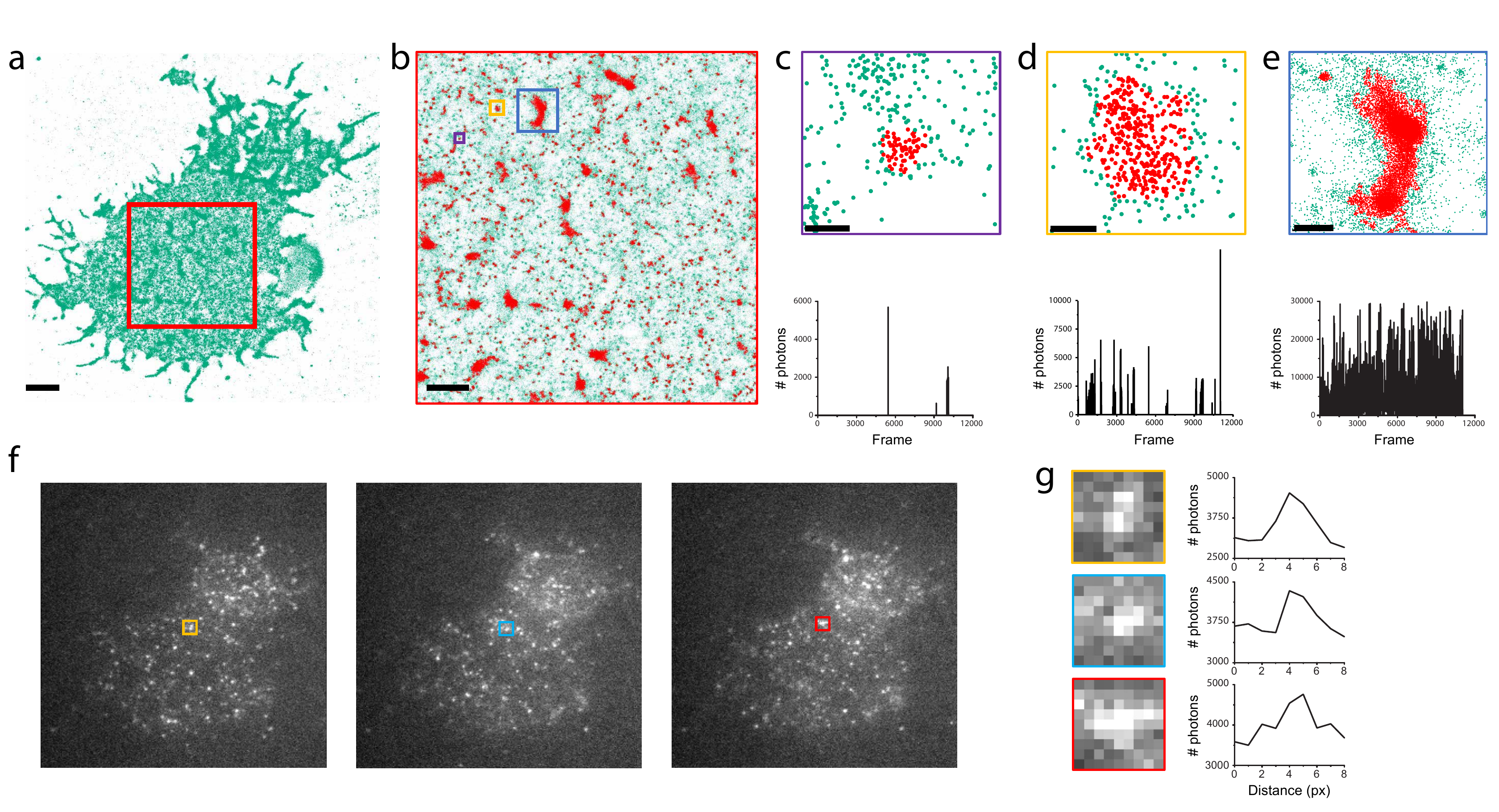}
	\caption{\\
		\\
		SMLM analysis of the plasma membrane organisation of CD4 variants.\\
		\\
		Plasma membrane organisation of the CD4 protein variants in resting T cells characterized using photo-activation localisation microscopy (PALM) followed by a Vorono\"{i}-based segmentation algorithm \cite{Chamma2016}. High density regions (HDRs) of irregular shape frequently forming networks of connected areas are identified. Yet, since the acquisition exhibited a high density of molecules with high blinking rates, the quantification of these HDRs can be affected by localization errors and under- or overcounting artifacts as described by Burgert et al. \cite{Burgert2015}. 
		(a) Original dataset composed of 1,747,681 localizations, scale bar 2 $\mu$m. (b) Magnification of the central area of the cell. Segmented HDRs are displayed in red, scale bar 1 $\mu$m. (c) Zoom on a HDR composed of 65 localizations, scale bar 50 nm.  As shown by its time trace, the localizations in this region are originating from a single fluorophore, making its blinking correction simple. (d-e) Zoom on a denser HDR (d, 384 localizations, scale bar 50 nm) and interconnected HDRs (e, 12,886 localizations, scale bar 200 nm) with their time trace. Even for a small HDR, the presence of multiple emitters complicates the blinking correction. The problem becomes even more apparent with interconnected HDRs because of their high-density of molecules. (f-g) Three frames of the original image acquisition, pixel size 105 nm. (g) Corresponding zoom to the region covered by one HDR. The high-density of molecules makes it difficult to properly separate each emitter, resulting in localization errors as well as under-counting.}	
	\label{fig:SMLMexamples} 
\end{figure} 

% Supplementary Fig. 7
\begin{figure}
	\centering\includegraphics[width=1\textwidth]{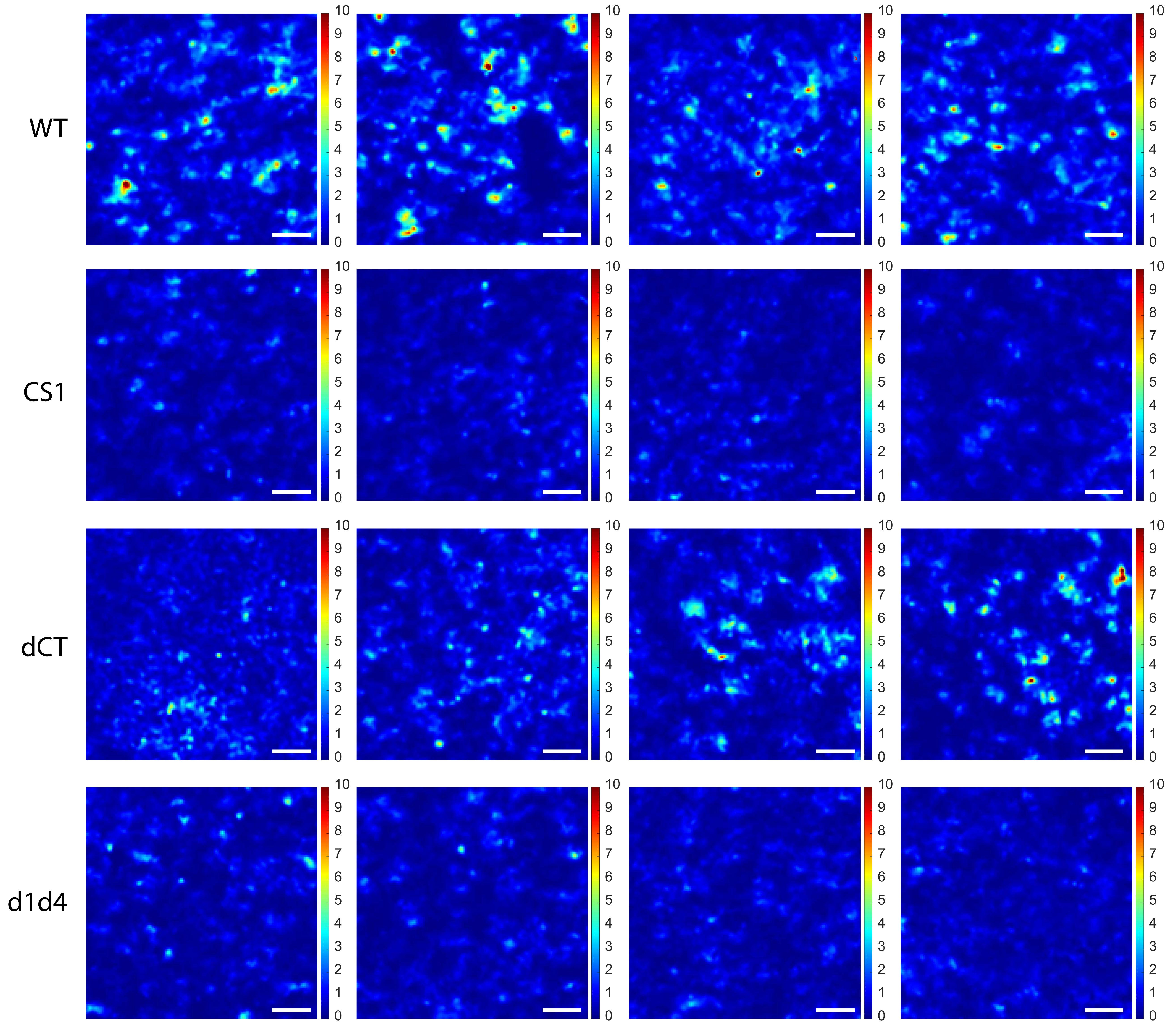}
	\caption{Each row represents four selected ROIs of one CD4 variant. Colorbar represents relative density ($ \# \delta_{avg} $). Scale bar 500 nm.}
	\label{fig:figure_selectionROIS2B} 
\end{figure} 

\FloatBarrier
\newpage
\section{Supplementary Note: SOFI density estimation}
The technical requirements for SOFI are a classical widefield microscope merged with a fast high sensitivity digital camera. SOFI image processing is based on higher order statistics and exploits the temporal sequence of blinking fluorescent emitters \cite{Dertinger2009,Dertinger2010}. Calculating spatio-temporal cross-cumulants allows SOFI to obtain a super-resolved, background-free and noise-reduced images.  Higher-order cumulants contain information about the photo-physics of the emitters. Combining SOFI images of different cumulant orders, allows one to extract physical parameters like molecular density \cite{Geissbuehler2012}, which we applied to investigate plasma membrane distribution of proteins.

\subsection{SOFI principle and theory}

 As stated by Dertinger et al. \cite{Dertinger2009}, the fluctuating emitters should switch between at least two optically distinguishable states (e.g. a dark and a bright state) repeatedly and independently in a stochastic manner. Images of stochastically blinking emitters are recorded such that the point-spread function (PSF) extends over several camera pixels. Acquiring a sequence of images results in a time dependent intensity trace for each pixel. Assuming N independently fluctuating emitters, the detected intensity is given as 
\begin{equation}\label{eq: observedIntensity}
	I(\mathbf{r},t)=\sum_{k=1}^{N}\epsilon_kU(\mathbf{r-r}_k)s_k(t) + b(\mathbf{r}) + n(\mathbf{r},t),
\end{equation} 
where $ \epsilon_k $ is the molecular brightness, $ U(\mathbf{r-r}_k) $ is the PSF at the position $ \mathbf{r}_k $, $ s_k(t) $ denotes a switching function (normalized fluctuation sequence, $ s_k(t) \in \{0,1\} $), $ b(\mathbf{r}) $ is a constant background, and $ n(\mathbf{r},t) $ represents an additive noise contribution.

For each pixel, an $n^{th}$ order cumulant is calculated for disentangling emitters inside the PSF. By applying the $n^{th}$ order cumulant to Eq. (\ref{eq: observedIntensity}), we obtain 
\begin{equation}\label{eq_cumulant_general}
	{\kappa{}}_n\{I(\mathbf{r},t)\}(\tau{})
	={\kappa{}}_n\left\lbrace\sum_{k=1}^M{\epsilon{}}_kU(\mathbf{r}-{\mathbf{r}}_k)s_k(t)+b(\mathbf{r})+ n(\mathbf{r},t)\right\rbrace(\tau{}).
\end{equation}
Using additivity and semi-invariance properties of cumulants \cite{Mendel1991}, the $n^{th}$ order cumulant with zero time lag can be written as
\begin{equation} \label{eq:nCumulant}
	\kappa_n\{I(\mathbf{r},t)\}=\sum_{k=1}^{N}\epsilon_k^nU^n(\mathbf{r-r}_k)\kappa_n\{s_k(t)\}
	+ \kappa_n\{b(\mathbf{r})\} + \kappa_n\{n(\mathbf{r},t)\}.
\end{equation} 
For ($ n \geq 2 $), the Gaussian noise ($ \kappa\{n(\mathbf{r},t)\} $)) and stationary background ($\kappa\{b(\mathbf{r})\} $ ) terms are eliminated by the cumulant analysis as an intrinsic property of cumulants. For an $n^{th}$ order cumulant, the PSF is raised to the  $n^{th}$ power (see Eq. \ref{eq:nCumulant}). As a consequence, the PSF is narrowed and the spatial resolution is improved by a factor of $\sqrt{n}$ \cite{Dertinger2009}. Therefore, increasing the cumulant order yields an image with an enhanced spatial resolution. Since a multiplication in the spatial domain corresponds to a convolution in the frequency domain, the cut-off frequency of the spectrum $\tilde{U}^{n}\left(\mathbf{k}\right)$ is n-times higher than that of $\tilde{U}\left(\mathbf{k}\right)$. By applying a deconvolution and a subsequent rescaling, the $n^{th}$ order cumulant image exhibits an up to n-fold resolution improvement \cite{Dertinger2010}. As shown in \cite{Dertinger2010}, virtual pixels can be calculated in between the physical pixels using cross-cumulants and followed by a flattening operation i.e. assigning proper weights to these virtual pixels \cite{Dertinger2010,Stein2015,Vandenberg2015a}. 

SOFI assumes a blinking model where the fluorophores reversibly switch between a bright and a dark state. In Deschout et al. \cite{Deschout2016a}, SOFI was applied to the PALM photo-physical model. In the PALM photo-physical model, the emitter activation is assumed as non-reversible, however, once the emitter is activated, it exhibits several fast blinking events prior to the final bleaching event \cite{Durisic2014}.
The emitter fluctuates between two different states (an on-state $ S_{\mathrm{on}} $ and a dark state $ S_{\mathrm{off}} $), which is expressed by the on-time ratio as
\begin{equation}
	\rho = \frac{\tau_\mathrm{on}}{\tau_\mathrm{on} + \tau_\mathrm{off}},
\end{equation}
where $ \tau_\mathrm{on} $ and $ \tau_{\mathrm{off}} $ are the characteristic lifetimes of the $ S_\mathrm{on} $ and $ S_{\mathrm{off}} $ states. The $n^{th}$ order cumulant $ \kappa_n\{s_k(t)\} $ is in this model described by a Bernoulli distribution with probability $ \rho_\mathrm{on} $ \cite{Geissbuehler2012} and approximated by an $n^{th}$ order polynomial function for the on-time ratio as
\begin{equation}\label{eq:cumulantFunction}
	f_n(\rho_\mathrm{on})=\rho_\mathrm{on}(1-\rho_\mathrm{on})\frac{\partial f_{n-1}}{\partial \rho_\mathrm{on}}.
\end{equation}

\noindent Under these conditions, the $n^{th}$ order cumulant can be approximated as \cite{Geissbuehler2012}
\begin{equation}\label{eq:approxCumulant}
	\kappa_n\{I(\mathbf{r,t})\} \approx \epsilon^n f_n(\rho_\mathrm{on})  \sum_{k=1}^{N}U^n(\mathbf{r-r}_k).   
\end{equation}

\subsection{Estimation of density maps}
 Geissbuehler et al. \cite{Geissbuehler2012} used three cumulant images ($2^\text{nd}$, $3^\text{rd}$, and $4^\text{th}$ order) to estimate molecular parameters: on-time ratio, brightness and molecular density. Here, we generalize this concept to any three cumulant images of distinct orders. If we assume spatially varying but locally constant on-time ratios and molecular brightness, the cumulants (for the cumulant order $ n > 1 $) can be approximated by \cite{Geissbuehler2012}

\begin{equation}
\label{eq:approxCumuln}
g_n(\mathbf{r}) \approx \epsilon^n(\mathbf{r}) f_n(\rho_\mathrm{on})N(\mathbf{r})   \mathcal{E}_V\{U^n(\mathbf{r})\}.
\end{equation}

\noindent where $ \mathcal{E}_V\{U^n(\mathbf{r})\} $ is the expectation value of $ U^n(\mathbf{r}) $, $ N(\mathbf{r}) $ is the number of emitters inside a detection volume V. Approximating the PSF near the interface in a total internal reflection (TIR) configuration by a lateral 2D Gaussian profile combined with an axial exponential profile, we obtain
\begin{equation}
\mathcal{E}_V\{U_\mathrm{TIR}^n(\mathbf{r})\} = \frac{c(\sigma_{x,y},\sigma_{z},d_z)}{n^{2}},
\end{equation}
where $ d_z $ represents the exponential decay of the TIR illumination \cite{Hassler2005}.

Using 3 consecutive cumulant images of orders $ n, (n-1), (n-2) $, we obtain for the ratios
\begin{align}
	\label{eq:comb234a}
	&K_1 = \frac{g_{n-1}}{g_{n-2}}  = 
	\frac{f_{n-1}(\rho_{\mathrm{on}}(\mathbf{r}))}{f_{n-2}(\rho_{\mathrm{on}}(\mathbf{r}))}
	\epsilon(\mathbf{r})
	\frac{\mu_{n-1}}{\mu_{n-2}}\\
	\label{eq:comb234b}
	&K_2 = \frac{g_{n}}{g_{n-2}}  = 
	\frac{f_n(\rho_{\mathrm{on}}(\mathbf{r}))}{f_{n-2}(\rho_{\mathrm{on}}(\mathbf{r}))}
	\epsilon^2(\mathbf{r})
	\frac{\mu_{n}}{\mu_{n-2}}\\
	\label{eq:comb234c}
	&K_3 = \frac{g_n}{g_{n-1}}  = 
	\frac{f_n(\rho_{\mathrm{on}}(\mathbf{r}))}{f_{n-1}(\rho_{\mathrm{on}}(\mathbf{r}))}
	\epsilon(\mathbf{r})
	\frac{\mu_{n}}{\mu_{n-1}},
\end{align}
where $ \mu_{n} = \mathcal{E}_V\{U^n(\mathbf{r})\} $. Substitution of Eq. (\ref{eq:comb234c}) into Eq. (\ref{eq:comb234a}) leads to

\begin{equation}
	\frac{g_ng_{n-2}}{g_{n-1}^2}  = 
	\frac{f_n(\rho_{\mathrm{on}}(\mathbf{r})) f_{n-2}(\rho_{\mathrm{on}}(\mathbf{r}))} {f_{n-1}^2(\rho_{\mathrm{on}}(\mathbf{r}))}.
\end{equation}

\begin{equation}
\left\{
\epsilon(\mathbf{r}) = \frac{g_n}{g_{n-1}} \frac{f_3(\rho_{\mathrm{on}}(\mathbf{r}))}{f_n(\rho_{\mathrm{on}}(\mathbf{r}))}
\frac{\mu_{n-1}}{\mu_{n}},
N(\mathbf{r}) = \frac{g_{n}(\mathbf{r})}{\epsilon^n(\mathbf{r})f_n(\rho_\mathrm{on})\mu_{n}}
\right\}.
\end{equation} 

Building up the ratios $ K_1 $ and $ K_2 $ from the Eq. (\ref{eq:cumul_params_K1}) for cumulants of $ 2^{nd},3^{rd},$ and $4^{th} $ order, we obtain 

\begin{align}
\label{eq:cumul_params_K1}
& K_1(\mathbf{r}) = \frac{\mu_{2}g_3}{\mu_{3}g_2}(\mathbf{r}) = \epsilon(\mathbf{r})(1 - 2\rho_\mathrm{on}(\mathbf{r}))\\
& K_2(\mathbf{r}) = \frac{\mu_{2}g_4}{\mu_{4}g_2}(\mathbf{r}) = \epsilon^2(\mathbf{r})
(1 - 6\rho_\mathrm{on}(\mathbf{r}) + 6\rho_\mathrm{on}^2(\mathbf{r})),
\end{align}

Solving for molecular brightness $ \epsilon $, on-time ratio $ \rho_{\mathrm{on}} $, we obtain two solutions for the on-time ratio $ \rho_{\mathrm{on}} $, and molecular brightness $ \epsilon $
\begin{equation}
	\label{eq:cumula_params_sol234}
	\left\{
	\rho_\mathrm{on}(\mathbf{r}) = \frac{3K_1^2 \pm K_1\sqrt{3K_1^2 - 2K_2} - 2K_2}{2(3K_1^2 - 2K_2)},
	\epsilon(\mathbf{r}) = \mp \sqrt{3K_1^2 - 2K_2}
	\right\},
\end{equation}
where the first solution corresponds to a negative brightness and will be discarded. The molecular density (number of molecules $ N $ per pixel area) is
\begin{align}
 N(\mathbf{r}) = \frac{g_2(\mathbf{r})}{\epsilon^2(\mathbf{r})\rho_{\mathrm{on}(\mathbf{r})}(1-\rho_{\mathrm{on}}(\mathbf{r}))}.
\label{eq:molparams_N}
\end{align}  

For cumulants of $ 3^{rd}, 4^{th}, $ and $5^{th} $ order, the ratios $ K_1$  and $ K_2 $ become
\begin{align}
	\label{eq:cumul_params_K1n345}
	& K_1(\mathbf{r}) = 
	\frac{
		\epsilon(\mathbf{r})
		(1 - 6\rho_{\mathrm{on}} + 6\rho_{\mathrm{on}}^2)
	}{
	(1 - 2\rho_\mathrm{on}(\mathbf{r}))
}\\
& K_2(\mathbf{r}) = \epsilon^2(\mathbf{r})
(12\rho_{\mathrm{on}}^2 -12\rho_{\mathrm{on}} + 1)
\end{align}
which ends in four solutions. Two correspond to positive molecular brightness
\begin{align}
	&\rho_\mathrm{on}(\mathbf{r})_{1,2} = 
	\frac{
		12K_1^3 \pm \sqrt{3}\sqrt{K_1^2(4K_1^2 - 3K_2) }\sqrt{4K_1^2 \mp 2\sqrt{K_1^2(4K_1^2 - 3K_2)} - 3K_2} -9K_1K_2
	}{6K_1(4K_1^2 - 3K_2)},\\
	&\epsilon(\mathbf{r})_{1,2} = \frac{
		\sqrt{4K_1^2 \mp 2\sqrt{K_1^2(4K_1^2 - 3K_2)} - 3K_2}
	}{\sqrt{3}}.
\end{align}

Using a combination of higher order cumulants for molecular parameters can theoretically provide higher spatial resolution of the molecular parameter maps assuming high enough SNR of the cumulant images used. For the combination of $ 4^{th}, 5^{th}, 6^{th}$ order cumulant, it is also possible to find a solution in a closed form, but due to its complexity, a numerical approach might be preferred. 

Therefore SOFI extracts density without counting individual events in the image. Density simply results from a correlation/cumulant analysis of intensity time traces.

%%-------------------------------------- Bibliography 
\newpage
\footnotesize{
	\bibliography{LukesSOFI2017}
	\bibliographystyle{ieeetr}}

\end{document}